\documentclass{artificial-life}
\usepackage[style=apa,natbib=true]{biblatex}
\AtEveryBibitem{
  \clearfield{howpublished}
  \clearfield{annotation}
  \clearfield{urlyear}
  \clearfield{urlmonth}
  \clearfield{urlday}
  \clearfield{urldateera}}

\usepackage{amsmath,amsfonts,amssymb}
\usepackage{overpic}

\newcommand{\dd}{\,\mathrm d}

\makeatletter
\renewcommand{\abs}[1]{\gdef\@abs{
    \begin{minipage}{0.8\textwidth}
        \singlespacing
        \textbf{Abstract. }#1
    \end{minipage}
    }}
\makeatother

\title{Continuous Game of Life:\\ cell emergence and self-organization at the edge of growth}

\auth{
    Alexandre Guillet \affil{1},
    Frank Jülicher \affil{1}
}

\corresponding{Alexandre Guillet}{aguillet@ik.me}

\affiliations{
    \item Max Planck Institute for the Physics of Complex Systems, Dresden, Germany
}

\abs{
    Conway's Game of Life shows that simple rules can generate a rich diversity of emerging structures.
    This cellular automaton has been translated to continuous space by~\citet{Rafler11} in a simulation called SmoothLife.
    The isotropic rule of this continuous Game of Life generates patterns whose beauty has attracted the attention of a growing community at the intersection of science and computer art.
    We study a minimal variant of this model, continuous in space and time, that generates cell-like patterns capable of self-replicating, gliding and disappearing.
    The phenomenology of these unit patterns is reported and related to homogeneous-state bifurcations, symmetry breaking, observed shape instabilities, finite-amplitude morphological changes, and a dilute-to-dense transition associated with cell proliferation.
    Its mapping onto a large reaction--diffusion system is interpreted in terms of homeostatic concentrations of morphogens, regulated by the nonlinear survival rule and generated through a cell-sourced cascade of auxiliary reactions.
    Introducing a global conservation law that limits resource availability causes the system to self-organize at this dilute-to-dense transition, which we call the edge of growth.
    A further exploration of parameter space reveals a variety of phases and the richness of life-like morphologies organized around this edge.
    Resemblance to biological processes such as division, motility, and death---together with a concise formulation and numerical implementation---makes the continuous Game of Life an appealing model system for investigating the emergence and self-organization of life-like patterns.
}

\keywords{continuous Game of Life, cellular automata, morphogenesis, self-organization, reaction--diffusion, phase transition}

\begin{document}

\coverpage

\section{Introduction}
\label{sec:intro}

\subsubsection*{Review}

Since its inception in 1970, the cellular automaton \emph{Game of Life} (GoL) defined by Conway~\citep{Gardner70}, famous for its complexity arising from simple rules on the 2-dimensional grid, has attracted sustained interest and inspired numerous developments.
After pioneering efforts in porting the GoL to continuous space~\citep{MacLennan90}, in particular its glider pattern via a scaling limit of large (inner and outer) discrete neighbourhoods~\citep{Evans01, Evans03, Pivato07}, the quest for more realistic Life-like automata has slowly faded from academia.

In contrast, the exploration of the \emph{Game of Life} became active in
online communities of enthusiasts\footnote{See \href{https://conwaylife.com}{conwaylife.com} and the Golly software.}
supported by the popularization of computers. This activity has produced increasingly complex structures within the original GoL rules, such as a \href{https://web.archive.org/web/20241216145426/https://conwaylife.com/wiki/8-bit_programmable_computer}{programmable computer} and a \href{https://web.archive.org/web/20260311171258/https://conwaylife.com/wiki/OTCA_metapixel}{meta-pixel} capable of emulating other cellular automata, illustrating its Turing-completeness~\citep{Rendell02}.
Despite extensive efforts, no finite \href{https://web.archive.org/web/20250328102558/https://conwaylife.com/wiki/Replicator#Examples_in_Life}{self-replicating machine} has been found in the GoL.

The continued interest in the GoL and its variants has resulted in the first successful implementation of a space-continuous isotropic rule, called \href{https://www.youtube.com/watch?v=KJe9H6qS82I}{\emph{SmoothLife}}~\citep{Rafler11}, based on disk and annulus-shaped integration domains in place of neighbour counting regions.
Simulations of gliders with better continuity in time have quickly followed, using an explicit Euler method modified with a clamp regularization~\citep{Rafler12, Hutton15}.
This breakthrough has laid the foundation for a renewed exploration of Life-like continuous systems inspired by cellular automata but freed from discrete states and the 4-fold symmetry of the grid. We call them \emph{continuous Games of Life} (cGoL).

Among \emph{SmoothLife} variations, several conceptual simplifications were introduced while maintaining a rich phenomenology.
Most relevant for this article, the little-known \href{https://web.archive.org/web/20230611103712/https://www.shadertoy.com/view/XtVXzV}{\emph{SmootherLife}} refinement from CornusAmmonis (2015) unifies the inner and outer integrals as convolutions with respect to a small kernel and a large kernel with the same Gaussian shape~\citep{CornusAmmonis17,Hutton15}.
This model generates clear self-replicating unit patterns made of a nucleus and a shell that resemble dividing (and sometimes dying) biological cells. This striking observation has motivated our investigation.

Another variation consists in reducing the nonlinearity of the GoL's survival rule, from its original bivariate form to a univariate function of the convolution integral. Called \emph{Lenia}, this simplification comes at the cost of a more sophisticated shape for the unique isotropic convolution kernel~\citep{Chan19}. The resulting simulation can generate single gliders (or oscillators) with surprisingly diverse morphologies, leading to the proposal of a naturalistic classification for artificial lifeforms. This exploration is pursued in~\citep{Chan20}, where the extended \emph{Lenia} model returns to a multivariate survival function (restricted to a sum of univariate ones), along with other higher-dimensional extensions.

Further work on \emph{Lenia} has identified several limitations and motivated extensions.
In spite of the small time steps, discreteness in time still plays an important role in the stability of the patterns: \emph{Lenia}'s gliders tend to be destabilized and vanish in the continuous-time limit~\citep{Davis22,Kojima23}. When reformulating the \emph{Lenia} model without the clamp regularization step, gliders can still be found in the continuous-time limit although with differing morphologies~\citep{Kawaguchi21}. The mathematical implications of these two modelling approaches are discussed in~\citep{Calcaterra23}; \citet{Yevenko25} analyse continuous-time formulations with dynamical-systems tools including symmetries, Lyapunov spectra, covariant Lyapunov vectors, and attractor dimensions.
The resulting integro-differential equation can be interpreted in terms of a particular reaction--diffusion system~\citep{Kojima23}: the slow field reacts to a combination of many fast-diffusing and -decaying species that it generates.

Furthermore, \emph{Lenia}'s gliders are metastable local structures, coexisting with stable global solutions: they readily vanish or explode into a space-filling pattern upon interaction. To address this issue, an environmental feedback from resource rarefaction has been introduced in~\citep{Suzuki23}.
Finally, a family of models enforcing a strict local conservation law on extended \emph{Lenia} has been introduced: \emph{Flow-Lenia}~\citep{Plantec23} proposes a complete reformulation in terms of fluxes, and \emph{MaCELenia}~\citep{Papadopoulos25} refines it into a compact continuity equation.

In this article, we combine these modelling approaches into a continuous and minimal formulation of the \emph{Game of Life} that produces clear self-replicating and motile unit patterns.

\subsubsection*{Roadmap}

We first introduce the cGoL model, moving from a general formulation of the dynamics to a parsimonious parametrization with 6 + 1 parameters. In two dimensions, a fine-tuned reference choice generates cell-like patterns that divide, glide and oscillate, or disappear, while nearby parameter values reveal a broader diversity of behaviours and morphologies. This phenomenology is charted empirically, from the homogeneous states and their saddle-node structure, through observed symmetry breaking, shape instabilities and finite-amplitude morphological transitions, occurring in the vicinity of a collective dilute-to-dense phase transition.

We then cast the same equations in reaction--diffusion terms: auxiliary fields become fast-relaxed morphogen concentrations, held within homeostatic ranges set by the nonlinear survival rule, while the Gaussian shape of the kernels emerges from an underlying cascade of auxiliary reactions. We next introduce a global constraint on a finite resource, consumed by growth and released by decay, and derive it as the well-mixed limit of a locally mass-conserving dynamics. For a sufficient volume, the resulting resource feedback drives a system initialized in the dense phase to self-organize at the edge of growth, the boundary with the dilute phase. Finally, scanning a targeted slice of parameter space maps the phase structure around this edge, where morphologies are richest and most life-like.

The article concludes with a discussion of dynamical-systems perspectives, self-organization and criticality, and questions of dimensionality and evolution.

\section{Definition of a continuous Game of Life}
\label{sec:definition}

The cGoL transposes the GoL into the continuum: neighbour counts become spatial convolutions, and the binary update table becomes a smooth target function. We first state the general dynamics, then specialize to the parametrization used throughout this article.

\subsection{General dynamics}

Let $L(\mathbf{x},t)$ be the ``Life''
field of a \emph{continuous Game of Life} (cGoL), in Euclidean space $\mathbf{x}\in \mathbb{R}^d$ and time $t>0$, taking scalar values in $\mathbb{R}$.
The continuous dynamics is determined by a nonlinear partial integro-differential equation of the form
\begin{align}\label{eq:cGoL}
    \partial_t L &= \Gamma(\{\Phi_k \ast L\})-L \\
    \Phi_k \ast L(\mathbf{x},t) &=\int \Phi_k(\mathbf{x}-\mathbf{y})L(\mathbf{y},t) \dd \mathbf{y} \quad .\label{eq:conv}
\end{align}
Here, $\Gamma$ is a nonlinear function of multiple variables indexed by $k=1,2,\dots$, extending the univariate (continuous-time \emph{Lenia}) case formulated in~\citep{Kawaguchi21}.
The variables $\Phi_k \ast L$ are auxiliary fields: smoothed versions of $L$ that represent neighbourhood population states.
The convolution integral Eq.~\eqref{eq:conv} is the continuous analogue of neighbour counting: it locally averages $L$ around any spatial coordinate $\mathbf{x}$.

Convolution kernels $\Phi_k$ correspond to different types of neighbourhood, typically a small one ($k=1$) and a large one ($k=2$) about 3 times wider. Isotropic kernels, $\Phi_k(\mathbf{x})=\Phi_k(|\mathbf{x}|)$, are a natural choice to restore the symmetry lacking in the discrete model. A constant kernel amounts to a simple integration, and will be introduced through a global conservation law ($k=3$).

The field $\Gamma(\{\Phi_k \ast L\})$, evaluated from neighbourhood states at each point in space, is called the \emph{target}. It can change as the primary field $L$ relaxes locally towards it, with the relaxation timescale taken as the time unit in Eq.~\eqref{eq:cGoL}. Defining the target field of the \emph{continuous Game of Life} amounts to specifying the discrete rule in the original \emph{Game of Life}.

\begin{figure*}[!htb]
    \centering
    \includegraphics[width=\linewidth]{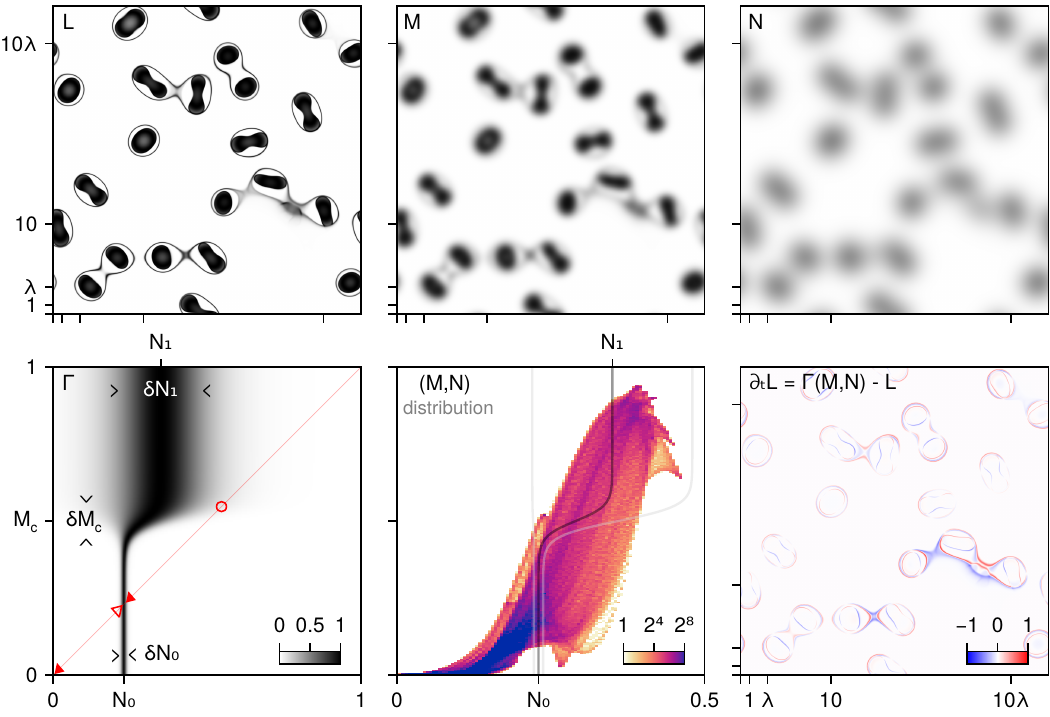}
    \caption{
        Primary and auxiliary fields: $L$, $M=\Phi_1\ast L$ and $N=\Phi_2\ast L$ (top, from left to right). Target function $\Gamma$ (bottom left, same grey scale as for $L, M, N$). Special homogeneous states are marked in red: stable and unstable post-bifurcation steady states (full and empty triangles), and pre-bifurcation non-steady optimum (empty circle). Spatial distribution of pairs $(M,N)$, where $\Gamma$ gets evaluated (bottom centre, pixel count). Growth or decay rate of the field $L$, Eq.~\eqref{eq:cGoL} (bottom right).
    }
    \label{fig:fields}
\end{figure*}

\subsection{Specific rule}

In this article, we remain close to the original rule of the \emph{Game of Life} in two ways: by starting with a bivariate target function (survival rule), as in \emph{SmoothLife}, and by using the same shape for the corresponding convolution kernels (neighbourhoods) as in \emph{SmootherLife}. By adapting those precursor models to the dynamics Eq.~\eqref{eq:cGoL}, we complete the continuous formulation of the GoL.

Instead of discrete squares (of widths 1 and 3), the Gaussian shape describes neighbourhoods as isotropic and smooth kernels:
\begin{equation}
\begin{aligned}\label{eq:kernel}
    \Phi_1(\mathbf{x})&=e^{-\pi|\mathbf{x}|^2} \quad , \qquad &&M=\Phi_1\ast L \\
    \Phi_2(\mathbf{x})&=\lambda^{-d}\Phi_1(\mathbf{x}/\lambda)  \quad , \qquad &&\,N=\Phi_2\ast L \quad .
\end{aligned}
\end{equation}
The two auxiliary fields $M$ and $N$ are thus Gaussian blurs of the primary field, averaging it at a small and a large neighbourhood scale, respectively, as illustrated in Fig.~\ref{fig:fields} (top). They play the role of continuous inner and outer neighbour counts, while the target function $\Gamma(M,N)$ specifies which combinations of these counts favour growth. We use the small scale as the unit length and set the large-to-small scale ratio to its original value $\lambda = 3$.
A reaction--diffusion interpretation is discussed in Section~\ref{sec:reaction-diffusion}.

By evaluating the target function $\Gamma$ for pairs of values $(M,N)$ at every spatial coordinate, one obtains the target field $\Gamma(M,N)$ that drives the dynamics Eq.~\eqref{eq:cGoL} of the primary field $L$, as illustrated in Fig.~\ref{fig:fields} (bottom).
The target is chosen to take values in $[0,\,1]$, so that this interval is absorbing for $L$: values initialized or perturbed outside it relax towards it.
The function $\Gamma$, replacing the Boolean update table of the GoL, specifies which large-neighbourhood populations $N$ favour growth; the centre $N_c(M)$ and width $\delta N_c(M)$ of this survival range are modulated by the small-neighbourhood population $M$.
Building upon previous definitions that parametrize the kink of the discrete rule, we retain the form represented in Figs.~\ref{fig:fields} and~\ref{fig:morpho},
\begin{align}\label{eq:rule-param}
    \Gamma(M,N) &= S'(\tfrac{N-N_c(M)}{\delta N_c(M)}) \\
    N_c(M) &= N_0 + (N_1-N_0)S(\tfrac{M-M_\text{c}}{\delta M_\text{c}})  \quad , \qquad S(m) = (1+e^{-4m})^{-1}    \label{eq:sigmoid} \\
    \delta N_c(M) &= \delta N_0 + (\delta N_1-\delta N_0)S(\tfrac{M-M_\text{c}}{\delta M_\text{c}}) \quad , \notag
\end{align}
where the derivative $S'$ of the sigmoid is a bell shape: $S'(n)=4S(n)(1-S(n))=\text{sech}(2n)^2$. The target parameters are specific values of the auxiliary fields $M$ and $N$ denoted with an index, see Eq.~\eqref{eq:ref}. This definition of the survival range effectively reduces parametric degrees of freedom by one as compared to~\citep{Rafler11,CornusAmmonis17}. The locality of the growth region appears crucial; other fast-decaying or compact bell shapes such as the Gaussian or the bump functions yield comparable results.

The equivalent form
\begin{equation}\label{eq:rule-param-bis}
    \Gamma(M,N) = S'(n_0+(n_1-n_0)S(\tilde m)) \quad ,\qquad
    (\tilde m,n_0,n_1) = (\tfrac{M-\tilde M_\text{c}}{\delta M_\text{c}},\tfrac{N-N_0}{\delta N_0},\tfrac{N-N_1}{\delta N_1})
\end{equation}
arises from properties of the sigmoid function $S$ defined in Eq.~\eqref{eq:sigmoid}. Note that the first parameter differs from the corresponding one in Eq.~\eqref{eq:rule-param}: $\tilde M_\text{c}=M_\text{c}-\tfrac{\delta M_\text{c}}{4}\log \tfrac{\delta N_1}{\delta N_0}$.

The rule of the Game is fully specified by the shapes $\Gamma$ (based on the sigmoid $S$) and $\Phi$ (Gaussian), and their parameters $(M_\text{c},\delta M_\text{c}, N_0, \delta N_0, N_1, \delta N_1)$ and $\lambda$. Given these parametrizations of the target function and kernels, the parameter space thus has 6 + 1 dimensions (to which one can add the dimensionality of space $d$).
In this article, the reference system uses Eqs.~(\ref{eq:rule-param}--\ref{eq:sigmoid}) in space with $d=2$ dimensions, original scaling $\lambda=3$, and target parameters
\begin{equation}\label{eq:ref}
    \mathbf{p}=(M_\text{c},\;\, \delta M_\text{c},\;\, N_0,\;\, \delta N_0,\;\, N_1,\;\, \delta N_1) =(0.50,\, 0.10,\, 0.23,\, 0.015,\, 0.35,\, 0.26) \quad .
\end{equation}
These parameters have been fine-tuned (and rounded to two significant digits),
so that the elementary localized state can divide several times, glide while oscillating, and disappear.
Because of this evocative phenomenology, illustrated in Fig.~\ref{fig:morpho} and detailed in the next section, we will refer to this emergent unit pattern as a cell.

\section{Emergence of cell-like patterns}
\label{sec:emergence}

We now turn from the rule itself to the structures it generates. Near the reference parameters Eq.~\eqref{eq:ref}, localized states emerge as coherent units with a nucleus and a shell, able to divide, move, oscillate, or disappear depending on their symmetry, perturbations and interactions. We first examine the homogeneous states, which anchor this phenomenology, before describing the resulting morphologies and shape transitions, which eventually hint at the proximity of a collective dilute-to-dense phase transition.

\subsection{Homogeneous states and bifurcations}

Some aspects of this system can be discussed from the scalar case $d=0$, which corresponds to homogeneous solutions for $d>0$. In this case, the convolution kernel is just a coefficient, equal to one since normalized in Eq.~\eqref{eq:conv}: $L=M=N$.
Near the reference parameters Eq.~\eqref{eq:ref}, the system has three homogeneous steady states (fixed points), marked with triangles in Fig.~\ref{fig:fields}.
These solutions of the equation $\Gamma(L_\ast,L_\ast)=L_\ast$ are practically indistinguishable from the solutions of $S'(\tfrac{L_\ast-N_0}{\delta N_0})=L_\ast$ as they occur in the lower region of the target before the transition ($N_0 < M_\text{c}-2\delta M_\text{c}$).

Since the survival range in this region is very thin, $\delta N_0 \ll N_0$, the lowest (empty) homogeneous steady state is practically vanishing, and it is stable (\emph{Game over}). The two other ones lie in the vicinity of $N_0 \pm \delta N_0$, hence they are very close to one another and related to the shell of the cell pattern. This pair of unstable and stable fixed points may be understood as reflecting the proximity of a saddle-node bifurcation.
An empty state with exact zero value, or a crossing of the bifurcation would both require a reparametrization of $\Gamma$. The two stable homogeneous states are especially relevant and denoted $L_\ast^{\pm}$.

For widely different target parameters, another saddle-node bifurcation can occur. The local maximum of $\Gamma(L,L)-L$ corresponds to the non-steady homogeneous state closest to the bifurcation. This special value, marked with a circle in Fig.~\ref{fig:fields}, depends on the four other parameters, $M_\text{c}, \delta M_\text{c}, N_1, \delta N_1$, and relates to the nucleus of the cell pattern.

The parrot-shaped distribution of joint values $(M,N)$ shown in Fig.~\ref{fig:fields} (bottom centre), which corresponds to an inhomogeneous non-steady state $L$, has visible traces of these special values as cusps and a recess near their location.

The rich morphogenesis generated by this nonlinear system suggests many further transitions between localized states, which lie well beyond the homogeneous-state analysis above. We therefore investigate these phenomena empirically.

\begin{figure}[ht]
    \centering
    \includegraphics[width=\linewidth, trim=5pt 0 5pt 0,clip]{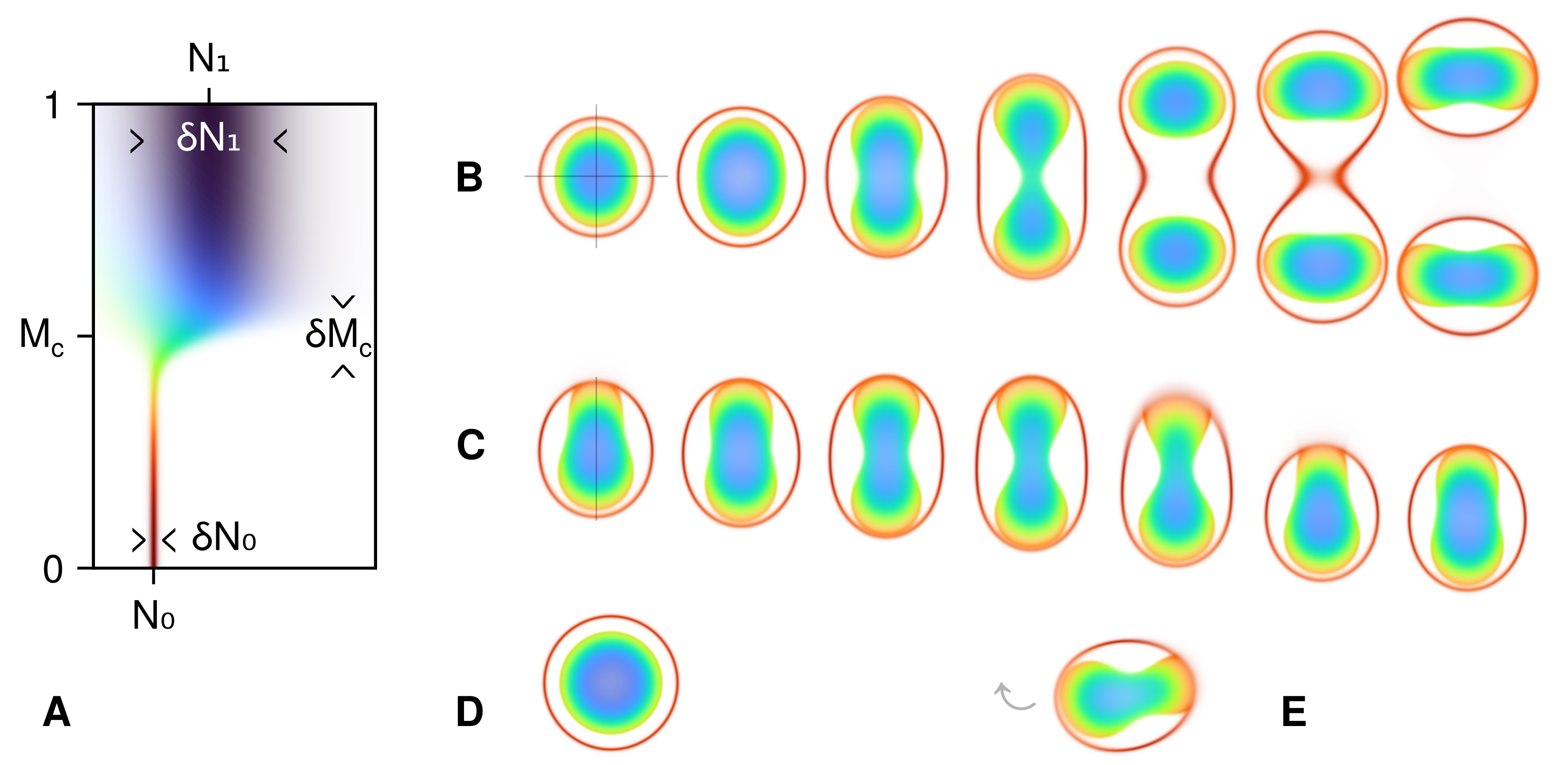}
    \caption{
        Cell dynamics at and near reference parameters Eq.~\eqref{eq:ref}, with colour coding of corresponding regions of the target function $\Gamma$ (A) and the field $L$ (B--E). Reference behaviours: dividing cell (B) or oscillating glider (C), depending on symmetry axes (left). Time advances from left to right in steps of 10 (B) and 6 (C) time units. (D--E) Variants: (steady) isotropic egg (D) and (clockwise) turning glider (E) for $N_1$ respectively increased and decreased by $0.03$.
    }
    \label{fig:morpho}
\end{figure}

\subsection{Morphological diversity and transitions}

Near the reference parameters Eq.~\eqref{eq:ref}, the \emph{continuous Game of Life} has a diversity of steady and non-steady localized states, all variants of a recognizable elementary pattern.
Activated regions $L$ form a nucleus and a shell, which correspond respectively to the broad and narrow parts of the survival region in the target function $\Gamma$, for high and low $M$. This correspondence is colour-coded in Fig.~\ref{fig:morpho}. The inner and outer empty areas correspond to the regions to the right and to the left of the survival range respectively.
By analogy with biology, we call this unit pattern a \emph{cell}, due to its self-replication, motility and disappearance.

In unfavourable regions of the parameter space where the unit pattern can still be initialized\footnote{The seed $L(0,\mathbf{x})=e^{-\pi \mathbf{x}^T \Sigma^{-1} \mathbf{x}}$ with a slight elongation facilitates the first division by breaking isotropy.} without vanishing, the cell remains in the form of a dormant \emph{egg}, an isotropic localized steady state shown in Fig.~\ref{fig:morpho} (D). Bistability can already be observed, with two possible radii of the egg at certain parameter values.

For more favourable parameters, the egg grows and can undergo various shape instabilities, triggered by environmental perturbations (such as interaction with neighbours).
The first observed symmetry breaking polarizes the egg, defining long and short symmetry axes.
The elongation of the cell can lead to its \emph{division}, while keeping its two symmetry axes (see Fig.~\ref{fig:morpho} B).
Some parameters lead to \emph{failed division}, in which case the cells die shortly after the division of the nucleus. If isotropy is not broken, the egg may grow into a \emph{bubble}.

The short symmetry axis can become unstable, leading to a second symmetry breaking. The result is a \emph{glider}: a bilateral cell morphology associated with motion. The glider can have a variety of behaviours and shape instabilities.
In the reference case illustrated in Fig.~\ref{fig:morpho} (B, C), the short axis is metastable, so that dividing cells can be destabilized into a glider. Conversely, interacting gliders can stop and divide.
These switches are changes of dynamical regime triggered by interactions, rather than linear instabilities of an isolated pattern.
This glider is notable for its shape \emph{oscillations} during motion.
When amplified, this oscillation leads back to cell division, with intermediate cases for which division fails asymmetrically. For other parameters, oscillations can be damped, resulting in a glider with a steady shape.

A further symmetry breaking of the long axis typically yields a clockwise or anti-clockwise \emph{turning glider}, shown in Fig.~\ref{fig:morpho} (E), with various radii for its circular trajectory. A \emph{tripod}, a steady cell with order 3 symmetry, can also be found, and observed to switch its morphology to different gliders upon perturbation.

The shape instability of an elongating cell may also be absent or insufficient to produce a successful division. This situation yields various extended patterns.
If the division of the nucleus is not followed by the scission of the shell, an intermediary body can form in the centre, which may turn into a full nucleus. Otherwise, segmented \emph{chains} or non-segmented \emph{filaments} can grow and sometimes percolate. A nucleus that elongates but has difficulty dividing leads to a \emph{snake}, with growing or shrinking ends. An ever-growing snake can fill space and generate various mazes and Turing patterns.

The phenomenology near the reference parameters is unexpectedly diverse, in spite of the simple morphology of the cell-like pattern compared with many \emph{Lenia} gliders.
Taken together, these observations suggest that the reference cell is poised among several long-lived localized morphologies, connected by symmetry breaking, shape instabilities, and finite-amplitude transitions.

\subsection{Dynamical phase transition}

The localized transitions described above motivate a broader question: how close are the reference parameters to a collective transition?
To make this question tractable in the 6-dimensional target-parameter space, we first probe a single control direction, then visualize selected additional directions through spatial parameter gradients.

Consider the scaling of the target function $r\Gamma$ with a constant coefficient $r$ (close to 1): $r>1$ enhances the growth rate whereas $r<1$ reduces it, according to Eq.~\eqref{eq:cGoL}. As a result, the field $L$ takes values in the interval $[0,\, r]$.
Since the scaling coefficient $r$ is constant, this modulation of growth amounts to scaling all auxiliary fields, by redefining $L$ on $[0,\, 1]$.
In turn, this is equivalent to rescaling all target parameters, given the parametrization Eq.~\eqref{eq:rule-param} of $\Gamma$. One obtains a simple interpretation of a perturbation along a particular direction of the parameter space:
\begin{equation}\label{eq:perturbation}
     r \Gamma(M, N;\mathbf{p}) \sim \Gamma(r M,r N;\mathbf{p}) = \Gamma(M,N;\mathbf{p}/r)
\end{equation}
where $\sim$ denotes the equivalence of the resulting dynamics up to a rescaling of $L$.
As expected, the scaling $\mathbf{p}/r$ of Eq.~\eqref{eq:ref} promotes growth for $r>1$ and inhibits it for $r<1$, interpolating between two coarse phases, dense or dilute, through many of the morphologies described above.

\begin{figure}[!ht]
    \centering
    \includegraphics[scale=1.2, trim=60pt 0 60pt 0, clip]{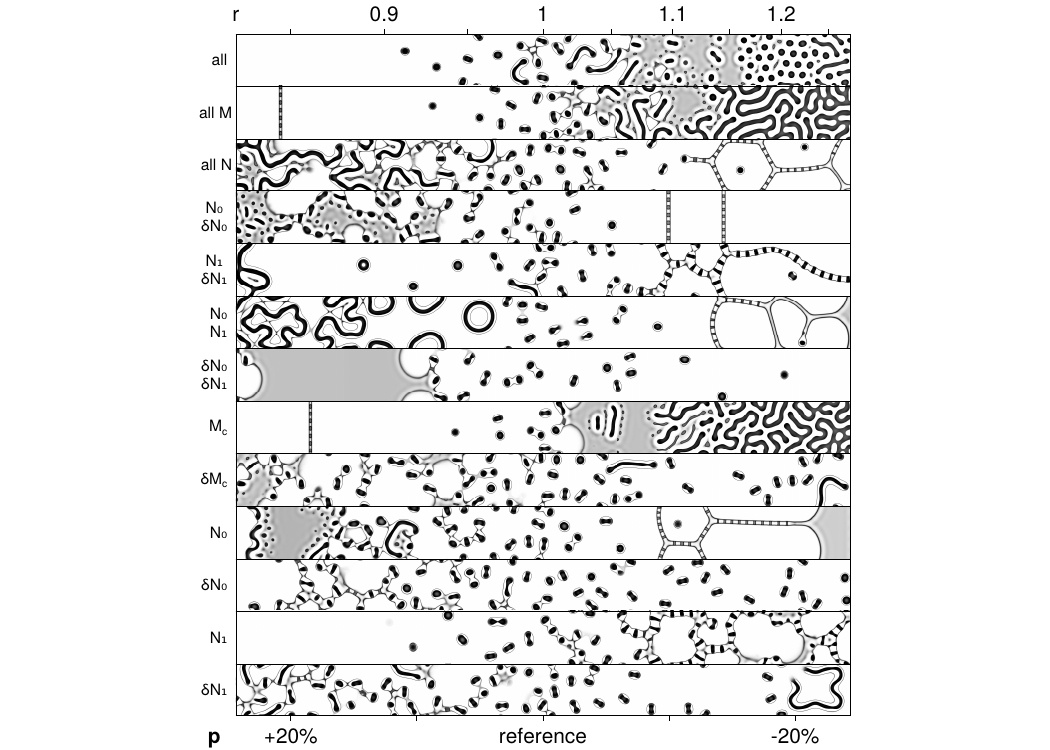}
    \caption{Dynamic states of the cGoL model in the presence of a spatial variation of the target parameters $\mathbf{p}=\mathbf{p}(\mathbf{x})$. For each panel, a subset of the parameters (specified on the left) is increased linearly from left to right around its reference value Eq.~\eqref{eq:ref} (centre). The top panel corresponds to Eq.~\eqref{eq:perturbation} with a position-dependent scaling $r=r(\mathbf{x})$ (top axis).
    }\label{fig:gradients}
\end{figure}

We further attempt to visualize the role of individual target parameters or subsets of them on the morphogenesis by simulating the cGoL with spatial parameter scans $\mathbf{p}=\mathbf{p}(\mathbf{x})$ along the horizontal direction. The resulting overview Fig.~\ref{fig:gradients} maps adjacent patterns, such as mazes, filaments and homogeneous states, around the reference cell and reveals the sensitivity of the tuning with respect to the different target parameters.

Fig.~\ref{fig:gradients} shows that the scaling Eq.~\eqref{eq:perturbation} of all target parameters (top panel) is dominated by the effect of scaling the auxiliary field $M$, in particular the first target parameter $M_\text{c}=\tfrac{1}{2}$. The isotropic egg and turning glider variants from Fig.~\ref{fig:morpho} (D, E), with values of $N_1$ differing by nearly $\pm 9\%$, can be mapped in Fig.~\ref{fig:gradients} in the second panel from the bottom. We note the propensity of this turning glider to assemble in chains.

These spatial scans suggest that the reference cell morphology has been selected close to a dynamical transition, but the evidence remains qualitative and restricted to $\lambda=3$.
Phase boundaries between dynamical regimes have likewise been identified in the parameter space of \emph{Lenia}-like systems~\citep{Papadopoulos24,Yevenko24,Hudcova26}.
After the reaction--diffusion reinterpretation of the next section, we return to this transition quantitatively: resource feedback identifies a transition value $r_\ast$ for $\lambda=3$, and a scan using $r$ and $\lambda$ as control parameters, with mean density as order parameter, maps the surrounding phase structure.

\section{Reaction--diffusion interpretation}
\label{sec:reaction-diffusion}

The nucleus and shell of the cell-like pattern occupy distinct regions of the target function evaluated from the auxiliary fields $M$ and $N$. We now reinterpret these equations as a coarse-grained reaction--diffusion system, in which the slow, non-diffusing field $L$ responds to fast-relaxed morphogen concentrations. The nonlinear reaction defines homeostatic morphogen ranges within the cell, while the Gaussian kernels arise from a cascade of auxiliary reactions.

\subsection{Morphogen concentrations at homeostasis}

As shown in~\citep{Kojima23} for continuous-time \emph{Lenia}, cGoL equations of the form Eq.~\eqref{eq:cGoL} can be mapped to the asymptotic regime of a reaction--diffusion system
\begin{equation}
    \tau_k \partial_t C_k =  \sigma_k^2 \nabla^2 C_k + f_k(\{C_j\})
\end{equation}
in which the primary field $C_0=L$ does not diffuse, $\sigma_0=0$, and multiple auxiliary concentration fields $C_k$ diffuse and react much faster, $\tau_k \ll \tau_0$ for $k\geq 1$.
The selected chemical species to which $L$ reacts are called \emph{morphogens}.
Eq.~\eqref{eq:cGoL} describes the slow nonlinear reaction $f_0=\Gamma(M,N)-L$, with morphogen concentrations $M$ and $N$.
Extended models in~\citep{Chan20,Plantec23} consider multiple primary (slow) fields.

The convolution form Eq.~\eqref{eq:conv} for morphogen fields is a coarse-grained description of the fast-relaxed subsystem, assumed to follow first-order kinetics: linear reactions $f_{k\geq 1}$ only allow conversion, decay and diffusion of auxiliary species.
Due to the timescale separation, each auxiliary field is linearly related to the slow field $L$ through a Green function that reflects the reaction network. When translation invariance is satisfied (with constant coefficients in unbounded or periodic space), these solutions reduce to convolutions: $C_k = \Phi_k \ast L$.

The only nonlinearity is carried by the target function $\Gamma$ in the slow reaction.\footnote{Previous discrete-time formulations such as \emph{SmoothLife} or \emph{Lenia} also include $L$ in the nonlinearity through the clamp function used to saturate $L$ to $[0,1]$.}
Morphogen concentrations set the target value $\Gamma(M,N)$ towards which $L$ relaxes.
The survival rule encoded in $\Gamma$ therefore specifies the morphogen ranges in which $L$ is maintained or amplified:
for each value of $M$, growth is favoured only within a corresponding range of $N$, centred on $N_c(M)$ and of width $\delta N_c(M)$.
In the reference morphology, the shell lies in the low-$M$ region, where this range is narrow and close to the iso-concentration contour $N\simeq N_0$, whereas the nucleus occupies a broader region at higher morphogen concentrations.
This correspondence between morphology and concentration-space regions is colour-coded in Fig.~\ref{fig:morpho}.
Thus, the target parameters Eq.~\eqref{eq:ref} directly define \emph{homeostatic concentration ranges} for the morphogens within the emergent cell-like pattern.

\subsection{Fast reaction cascade and Gaussian kernels}

The Gaussian shape of the kernels Eq.~\eqref{eq:kernel} can be accounted for by a long cascade of fast auxiliary reactions. To see this, consider a chain of auxiliary species $C_k$, $k=1,2,\dots$, sourced by the slow field $C_0=L$, and described by the linear reaction--diffusion network
\begin{equation}\label{eq:cascade}
    \tau \partial_t C_k = \sigma^2 \nabla^2 C_k - C_k + C_{k-1} \quad.
\end{equation}
This simple network represents a cascade of diffusing and decaying species, each successively produced from the previous one and ultimately sourced by the slow field $L$.
Using the separation of timescales $\tau \ll 1$, relaxed states are formally expressed as
\begin{equation}
    C_k=[1-\sigma^2\nabla^2]^{-k} L = \phi_k \ast L \quad,
\end{equation}
where $\phi_k = \phi_1 \ast \dots \ast \phi_1$, obtained by Fourier inversion, is known as the Matérn kernel
\begin{align}\label{eq:Bessel-kernel}
    \phi_k(\mathbf{x}) &= \int \frac{e^{i 2\pi \mathbf{q}\cdot\mathbf{x}}}{(1+(2\pi \sigma |\mathbf{q}|)^2)^k}\dd \mathbf{q}
    = \frac{(|\mathbf{x}|/\sigma)^k K_{k-\frac{d}{2}}(|\mathbf{x}|/\sigma)}{2^{k-1}(k-1)!(2\pi \sigma |\mathbf{x}|)^\frac{d}{2}}
\end{align}
involving the modified Bessel function of the second kind, with exponential tail\linebreak $K_\nu(x)\sim\sqrt{\tfrac{\pi}{2x}}e^{-x}$. This kernel corresponds to the steady spatial profile of the $k^\text{th}$ species in the cascade, decaying and diffusing from a point source. This Green function is valid in unbounded space, $\mathbf{x}\in \mathbb{R}^d$. In the case of a short cascade, $k\leq \frac{d+1}{2}$, a singularity (or a cusp for $k>\frac{d}{2}$) is present at the origin.

The Gaussian kernels used in Eq.~\eqref{eq:kernel} are recovered in the limit of a long cascade $k\gg 1$: the $k$-fold convolution of the elementary Green function approaches a Gaussian by the central limit theorem, summarized at the level of the space operator as
\begin{equation}
    [1-\sigma^2\nabla^2]^{-k}\sim e^{k\sigma^2\nabla^2} \quad.
\end{equation}
The unit-scale kernel in Eq.~\eqref{eq:kernel} is matched by taking the scaling limit $\sigma^2\sim(4\pi k)^{-1}$.
The large- and small-neighbourhood morphogen fields $N$ and $M$ are thus idealized as two selected species in the $L$-sourced cascade Eq.~\eqref{eq:cascade}, located at depths $\lambda^2 k$ and $k\gg 1$.

Without a cascade, the simplest Bessel shape $\phi_1$ represents morphogens with different diffusion coefficients, directly sourced from $L$ without intermediate reactions. In this case, gliding does not occur and divisions are incomplete, with the systematic formation of an intermediary body. A small cascade with $M=\phi_1 \ast L$ and $N=\phi_9 \ast L$ (or as small as $\phi_4$) is sufficient for complete divisions.

In line with the mapping proposed by~\citet{Kojima23}, and extending its interpretation, the cGoL model can be viewed as a coarse-grained description of a large reaction--diffusion system in which the slow field responds nonlinearly to a few morphogens. Its distinctive feature is the non-standard nonlinearity encoded by $\Gamma$, inherited from the GoL survival rule, which directly specifies homeostatic morphogen concentration ranges and the associated growth target.
The Gaussian shape of the kernels arises in the long-cascade limit of a linear reaction network of fast-diffusing and -decaying auxiliary species.

\section{Resource limitation as self-tuning feedback}
\label{sec:resource}

Repeated cell division leads to proliferation limited only by domain size. To address this unphysical behaviour, we extend the model with a global coupling to a finite resource. This global constraint follows from a local conservation law, when the available resource diffuses fast. The associated feedback dynamically retunes the parameters, driving the system to the edge of growth between the dense and dilute phases.

\subsection{Global constraint from finite resource}

Assume that growth consumes resource from a finite reservoir, which is replenished when $L$ decays.
The fixed total resource $R$ is split into an available part $R_\text{a}$ and the mass stored in $L$:\footnote{Here we neglect the small nonzero background $L_\ast^-$. Strictly, on an unbounded domain one should either choose $\Gamma$ such that $L_\ast^-=0$, or define the consumed resource relative to the background as $\int (L-L_\ast^-)\dd\mathbf{x}$.}
\begin{equation}\label{eq:conservation}
    R =R_\text{a} +\int  L\dd \mathbf{x} \quad, \qquad
    \partial_t R = 0 \quad .
\end{equation}
Following the resource-feedback approach of~\citep{Suzuki23}, we define an abundance coefficient
\begin{equation}
    r(t) = \frac{R_\text{a}(t)}{R} \quad,
\end{equation}
which decreases from 1 to 0 as the resource is consumed. Inserting $r(t)\leq 1$ in the target function, as in Eq.~\eqref{eq:perturbation}, provides negative feedback on growth.

The temporal modulation $r(t)$ breaks the equivalence in Eq.~\eqref{eq:perturbation}, yielding two distinct feedback mechanisms. Type I limits growth by decreasing the target $r(t)\Gamma$, whereas type II decreases auxiliary fields, $r(t)M$ and $r(t)N$, or equivalently increases target parameters, $\mathbf{p}/r(t)$.
Type II can be visualized as pushing the system to the right in the top panel of Fig.~\ref{fig:gradients}, as the available resource is consumed. Both types introduce the mass $\int L \dd \mathbf{x} = 1 \ast L$, a third variable with constant kernel, in the target.

Figure~\ref{fig:gradients} indicates that the resource-limitation effect is dominated by the increase in $M_\text{c}$, motivating a third feedback that affects this parameter alone. Type III preserves the bivariate formulation of the target function $\Gamma(M,N)$ by perturbing the kernel for the first auxiliary field: $M = (\Phi_1 - \tfrac{M_\text{c}}{R}) \ast L$. The three feedback types are summarized as follows:
\begin{align}
    r(t)\Gamma(M,N;\mathbf{p}) & \qquad \text{type I} \nonumber \\
    \mathbf{p}(t)=\mathbf{p}/r(t) &\qquad \text{type II} \\
    M_\text{c}(t) = M_\text{c}(1+\tfrac{1\ast L}{R}) \simeq M_\text{c}/r(t) & \qquad \text{type III} \quad .\nonumber
\end{align}
The feedback mechanisms II and III dynamically and autonomously retune either all reference parameters Eq.~\eqref{eq:ref} or just the first one, with coupling strength $R^{-1}$.

\subsection{Local conservation and the well-mixed limit}

The global resource feedback can be viewed as the well-mixed limit of a locally conserved resource dynamics.
Let $\rho_\text{a}(\mathbf{x},t)$ denote the density of available resource, and $\rho$ the average density of total resource.
Feedback type I provides a natural form for the local dynamics:
\begin{equation}
\begin{aligned}
    \partial_t L &= r(\mathbf{x},t)\Gamma(M,N)-L \quad , \qquad r(\mathbf{x},t)=\frac{\rho_\text{a}(\mathbf{x},t)}{\rho}\\
    \partial_t(L+\rho_\text{a}) &= D_\text{a}\nabla^2\rho_\text{a} \quad .
\end{aligned}
\label{eq:local-resource}
\end{equation}
The total density $L+\rho_\text{a}$ follows a continuity equation with diffusive flux of the available resource, and therefore obeys a local conservation law.
The source terms in the dynamics of $L$ and $\rho_\text{a}$ are equal and opposite: growth of $L$ consumes locally available resource, whereas decay of $L$ replenishes it.
This environmental feedback constitutes a mass-conserving counterpart to the non-conserved version introduced in~\citep{Suzuki23}, where resource is locally consumed to maintain $L$, and continually restored to a target availability.

In the well-mixed limit $D_\text{a}\to\infty$, the available resource becomes homogeneously distributed, so the local abundance $r(\mathbf{x},t)$ reduces to $R_\text{a}(t)/R=r(t)$, recovering the global feedback. Spatial integration also recovers Eq.~\eqref{eq:conservation}, with $R_\text{a}=\int \rho_\text{a}\dd\mathbf{x}$.
The global resource model is thus a spatially averaged description, accurate when the available resource is fast-relaxing.

This finite-diffusion resource-feedback extension connects the original non-conserved cGoL dynamics to mass-conserving reaction--diffusion systems~\citep{Halatek18}.
Here $L$ is conserved only jointly with the resource reservoir, unlike in strictly conservative reformulations~\citep{Plantec23,Papadopoulos25}, but in line with their food-channel extensions.
The dynamics is generically non-variational because regulation by $M$ and $N$ lacks reciprocal thermodynamic back-reactions.\footnote{At fixed $\Gamma(M,N)$, the local conversion between $L$ and $\rho_\text{a}$ satisfies detailed balance and admits the free-energy density $L(\log L-1)+\rho_\text{a}(\log \rho_\text{a}-1) + L\log(\rho/\Gamma)$.}
Preliminary observations suggest that resource diffusion affects cell motion; we leave the analysis of this dynamics to future work.

\subsection{From volume-limited growth to resource-limited self-tuning}

We now quantify how global resource feedback controls the equilibrium state.
Given a finite total resource $R$ and a finite spatial domain $\Omega\subset\mathbb{R}^d$ of volume $|\Omega|$, we start from a low cell count to leave room for growth. Cells are counted as connected subsets of $\Omega$ satisfying $M>M_\text{c}$ (distinct nuclei).
Ensemble averages over equilibrated simulations provide estimates of the mean cell number $\nu$ and field density $\mu$, defined in Eq.~\eqref{eq:order-param}; the corresponding mass per cell is $m=\frac{\mu |\Omega|}{\nu}$.

\begin{figure}[htb]
    \centering
    \includegraphics[width=\linewidth]{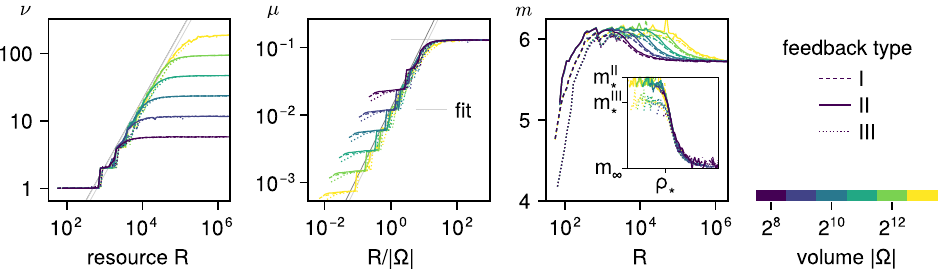}
    \caption{Equilibrium cell number $\nu$, average density $\mu$, and mass per cell $m$, as functions of resource $R$ and volume $|\Omega|$. Lines encode different volumes (colours) and resource feedback types: target scaling (I), morphogen scaling (II), and kernel shift (III). Grey lines show the asymptotes of the fitted Eq.~\eqref{eq:density}.
    }
    \label{fig:resource}
\end{figure}

Their dependence on the resource $R$ and volume $|\Omega|$ is summarized in Fig.~\ref{fig:resource}: $\nu$ scales with $|\Omega|$ in the volume-limited regime and with $R$ in the resource-limited regime; discrete steps remain visible at low cell numbers.
The crossover occurs near a total resource density $\rho_\ast$, and the average value of $L$ is well approximated by
\begin{equation}\label{eq:density}
   \mu \simeq\frac{\mu_\infty}{\sqrt{(\rho_\ast/\rho)^2 +1}} \quad , \qquad \rho = \frac{R}{|\Omega|} \quad .
\end{equation}
Fitting this model for $\nu\geq 4$ gives $\mu_\infty = 0.131 \pm 0.001$ for abundant resource (property of the reference cGoL model $r=1$), and $\mu\sim \mu_\infty\rho/\rho_\ast$ for limited resource, with feedback-dependent crossover densities $\rho^\text{(I, II)}_\ast = 10.5\pm 0.5$ and $\rho^\text{(III)}_\ast = 13.5\pm 0.5$.

Irrespective of the approximation Eq.~\eqref{eq:density}, types I and II converge to the same $\mu$ and hence to the same abundance $r=1-\mu/\rho$.
Equation~\eqref{eq:perturbation} then predicts identical cell morphologies, up to rescaling of $L$, giving $m^{(\mathrm{II})}=m^{(\mathrm{I})}/r$ and $\nu^\text{(II)}=r\nu^\text{(I)}$, visible in Fig.~\ref{fig:resource} and confirmed in its inset.

The mass per cell decreases from about $m_\ast^\text{(II)}=m_\ast^\text{(I)}/r_\ast=6.10\pm 0.05$ and $m_\ast^\text{(III)}=6.05\pm 0.05$ in the resource-limited regime to the volume-limited value $m_\infty=5.725 \pm 0.005$. At confluence, continued division and disappearance distribute cells across different elongation states; under resource limitation, division stops while all cells remain elongated with enhanced mass below the division threshold.
When sufficiently large, the average number of cells $\nu$ is
\begin{align}
    \nu \simeq \begin{cases}\displaystyle \frac{\mu_\infty R}{m_\ast\rho_\ast} \quad\;\,, \qquad \rho \ll \rho_\ast \\
    \displaystyle \frac{\mu_\infty |\Omega|}{m_\infty} \quad, \qquad \rho \gg \rho_\ast \end{cases}
\end{align}
with the specific relations between types I and II: $\rho_\ast^\text{(I)}=\rho_\ast^\text{(II)}$ and $m_\ast^\text{(I)}=r_\ast m_\ast^\text{(II)}$.

The equilibrium system with resource feedback effectively behaves as a retuned version of the model without feedback.
In the resource-limited regime $\rho\ll \rho_\ast$, always reached in the large-volume limit $|\Omega|\!\to\!\infty$ at fixed $R$,
the total mass approaches $\mu|\Omega|\!\sim \!\mu_\infty R/\rho_\ast$ rather than growing with volume.
During the transient growth, the system adjusts itself to a boundary point on the transition manifold between the extensive (volume-limited, dense) and non-extensive (resource-limited, dilute) phases of the parameter space.
Type II rescales all parameters to $\mathbf{p}/r_\ast$, with $r_\ast = 1-\mu_\infty/\rho^\text{(I, II)}_\ast \simeq 0.987$, whereas type III changes only $M_\text{c}$ to $M_\ast = (1+\mu_\infty/\rho^\text{(III)}_\ast)M_\text{c} \simeq 1.010\,M_\text{c}$.
Thus, the hand-tuned reference parameters of Eq.~\eqref{eq:ref} lie in the dense phase, within $\sim\!1$\% of the edge of growth, motivating a broader exploration of the surrounding phase structure.

\section{Exploration of phase structure}
\label{sec:phases}

The resource-feedback model identified an abundance value $r_\ast$ marking the edge of growth, for the neighbourhood scale ratio $\lambda=3$.
We now broaden this picture through an extensive scan of the field density and contrast across the $(r,\lambda)$ plane, mapping two background basins, each containing a stability boundary for single droplet patterns and an edge of growth. We then extend the self-tuning estimate of $r_\ast$ across $\lambda$, uncovering the richness of life-like morphologies that self-organize near this edge.

\subsection{Control and order parameters}

We first treat the resource abundance $r$ as a constant control parameter without feedback, implemented by moving along the parameter-space line $\mathbf{p}/r$, with $\mathbf{p}$ given by Eq.~\eqref{eq:ref}. By Eq.~\eqref{eq:perturbation}, this is dynamically equivalent to scaling the target function, while preserving the range $L\in[0,1]$. Increasing $r$ lowers the homeostatic concentration ranges and promotes growth, whereas decreasing $r$ inhibits it.

The neighbourhood scale ratio $\lambda$, previously fixed, controls the separation between the two averaging kernels and is the only geometric parameter of the model.
It therefore provides a natural second control parameter.

\begin{figure*}[htb]
    \centering
    \includegraphics[width=\linewidth]{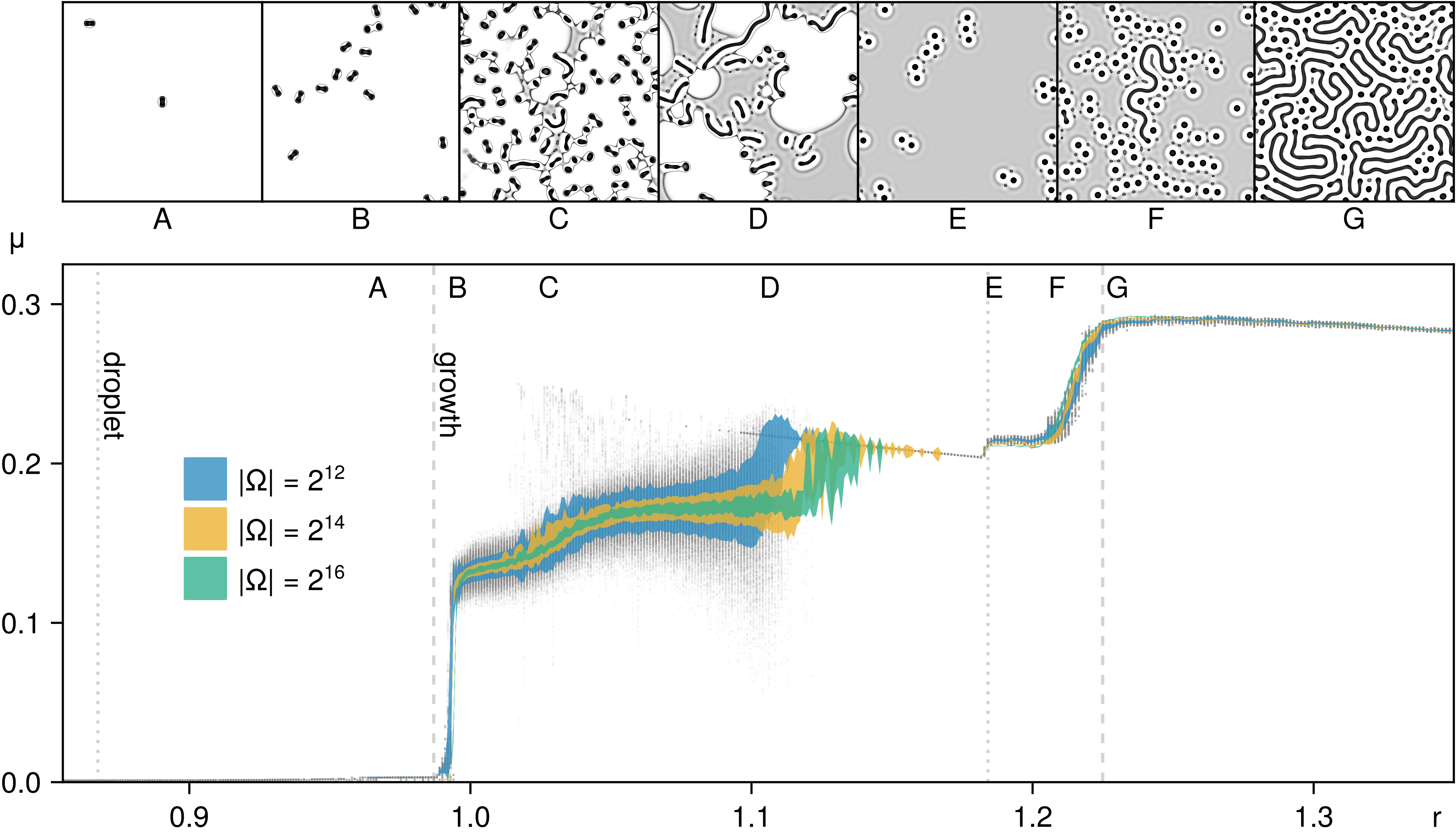}
    \caption{Scan of the density $\mu$ (order parameter), with constant resource abundance $r$ as control parameter. Standard deviation around the mean (coloured area) over independent simulations for a given volume $|\Omega|$. Droplet-stability boundaries and edges of growth (dotted and dashed lines). Scale ratio $\lambda=3$, initial state with 2 seeds, cutoff time $t_{\max}=5000$.}
    \label{fig:scan1}
\end{figure*}

The average density $\mu$, already studied for resource limitation, is a simple observable that we use as order parameter. In the absence of resource feedback, $\mu$ is
expected to jump near transition values of $r$, separating dilute from dense phases, in which growing patterns fill the available volume until the average density reaches its saturation value $\mu_\infty$.
This expectation is confirmed in Fig.~\ref{fig:scan1}, where several
transitions are visible in the profile $\mu(r)$ and later described; the sharpest cliff lies in the previously identified interval $r\in[0.987,1]$.

Density alone does not distinguish the upper homogeneous state (visible in Fig.~\ref{fig:scan1} between D and E) from inhomogeneous patterns. We therefore complement $\mu\in[0,1]$ with the
exponential entropy $\kappa\in[0,1]$ of the field-value distribution, a diversity index capturing the field contrast
\begin{equation}\label{eq:order-param}
\begin{aligned}
    \mu &= \int_0^1 \ell \mathbb{P}(\ell)\dd\ell \quad, \qquad \mathbb{P}(\ell) = \lim_{t\to\infty}|\Omega|^{-1}\int_\Omega \delta(\ell - L(\mathbf{x},t))\dd \mathbf{x}  \\
    \kappa &= \exp\left(-\int_0^1 \mathbb{P}(\ell) \log \mathbb{P}(\ell) \dd \ell\right) \quad.
\end{aligned}
\end{equation}
Here, $\log \kappa$ is the differential entropy of $\mathbb{P}(\ell)$ on the unit interval (span of $\Gamma$, hence of $L$). Equivalently, $-\log \kappa$ is its Kullback--Leibler divergence (relative entropy) from the uniform distribution $\mathbb{P}(\ell)=1$, which represents the most contrasted case $\kappa=1$. A homogeneous field $L=L_\ast$, $\mathbb{P}(\ell) = \delta(\ell-L_\ast)$, has vanishing contrast $\kappa = \!\dd \ell \to 0$ (histogram bin size).

Figure~\ref{fig:scan2} compares the density and contrast. Their variability across independent simulations serves as a diagnostic of slow convergence and sensitivity to initial conditions.

\subsection{Lower and upper background basins}

\begin{figure*}[!htb]
    \centering
    \includegraphics[width=\linewidth]{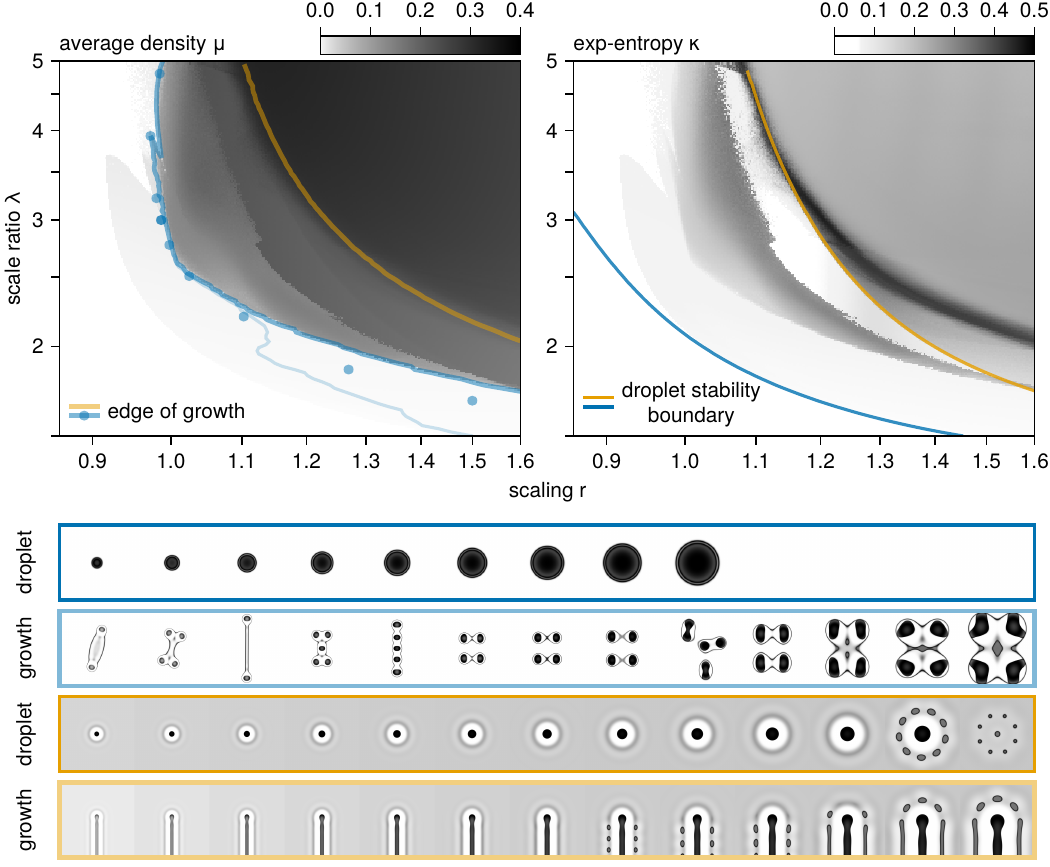}
    \caption{Scan of order parameters $(\mu, \kappa)$ with control parameters $(r,\lambda)$.
    The median over 20 trials is retained for $|\Omega|=2^{12}$, 5 $\lambda$-scaled seeds, $t_{\max}=1000$.
    Edge of growth (left, multiple estimates) and stability boundary for a single droplet (right) in the lower and upper background basins (blue and orange lines). Localized states along these lines (bottom, $\lambda$ increases from left to right).}
    \label{fig:scan2}
\end{figure*}

The numerous phases visible in Figs.~\ref{fig:scan1} and~\ref{fig:scan2} form a sequence that appears twice, see Fig.~\ref{fig:scan1} (A--C) and (E--G).
This repetition reflects the two stable homogeneous fixed points, $L_\ast^{-}$ and $L_\ast^{+}$, which can each serve as an empty background for localized patterns.
The changeover around Fig.~\ref{fig:scan1} (D) separates the corresponding background basins, at lower and higher $r$ respectively.

In Fig.~\ref{fig:scan1} ($\lambda=3$), the state $L_\ast^{-}\simeq 0$ is dominant for $r<1.1$, while $L_\ast^{+}\simeq N_0/r$ takes over for $r>1.1$, with overlap around $r=1.10\pm0.05$ in Fig.~\ref{fig:scan1} (C--D).
As $r$ increases, crowded cell or snake dynamical patterns on the $L_\ast^{-}$ background are increasingly bridged and surrounded by local patches of $L_\ast^{+}$, until chaotic fluctuations allow $L_\ast^{+}$ to percolate (D) and fully replace $L_\ast^{-}$.
When this happens, the localized patterns usually dissolve into $L_\ast^{+}$, which then becomes the empty background for patterns at higher $r$.

This changeover is broad and sensitive to simulation conditions: increasing the cutoff time lowers its apparent value of $r$, increasing the volume raises it (see Fig.~\ref{fig:scan1} D), and initial conditions also matter for $\lambda<3$. It appears in Fig.~\ref{fig:scan2} as a sharp but noisy drop in $\kappa$ that roughly separates the lower and upper background basins, whose inner boundaries are marked in blue and orange respectively.

\subsection{Boundary of droplet stability}

The first type of boundary delimits the stability of isolated droplets in the dilute regime.

The region of successful initialization, visible in Fig.~\ref{fig:scan2} at low $\mu$ and $\kappa$ and dependent on the seed characteristics, is more fundamentally bounded by the region of stability of localized states. The egg pattern in Fig.~\ref{fig:morpho} (D) is ubiquitous in the lower basin. A similar localized state exists in the upper basin in the form of a dot pattern. A single such droplet is only stable above the boundary lines plotted in each basin in Fig.~\ref{fig:scan2} (top right, also reported in Fig.~\ref{fig:scan1}, dotted lines).

These stability boundaries are determined accurately, albeit numerically, by bisecting $r$ starting from successful seeds (whose radii scale with $\lambda$).
They do not constitute an absolute bound for the existence of stable inhomogeneous states: droplets in dimension $d=1$, embedded in $d=2$ as a stripe pattern, can be generated at lower $r$ from a space-spanning initial seed.
Other localized structures can persist below the upper dot-stability boundary, particularly for $\lambda>3$, see Fig.~\ref{fig:scan2} (top right, orange line). These include bubbles and residual small-scale patterns that survive after the central dot vanishes.

At the limit of survival of an egg pattern (lower-$r$ droplet), the gap between the shell and the nucleus vanishes, as illustrated in Fig.~\ref{fig:scan2}. Collective effects can extend the survival region of these droplets, echoing the original rule of the \emph{Game of Life}.

\subsection{Edge of growth: collective dilute-to-dense transition}

The second type of boundary marks the transition between a dilute phase, with eggs (or non-growing cells) and dots (Fig.~\ref{fig:scan1} A and E, respectively), and a dense, space-filling phase (C and G). The lower dense phase (C) is highly dynamic, with dividing-dying cells and growing-shrinking snakes, whereas the upper one reaches a static, space-filling configuration (the Turing-like pattern in G).

In both basins, values of $\mu$ and $\kappa$ in the dilute phase depend on the initial state and the number of surviving patterns (with $\mu\geq L_\ast^{\pm}$), whereas they approach characteristic saturation values in the dense phase.

For $\lambda=3$, the lower edge of growth was previously estimated at $r_\ast\simeq0.987$ from the self-tuning of $r$ by resource-limitation feedback at large volume.
At constant $r$, however, growth can remain difficult to ignite slightly above the edge, because spontaneous growth from a single isolated cell often requires $r>r_\ast$. A simulation initialized with a single seed can therefore remain dilute, although growth may be triggered by a finite-amplitude perturbation.\footnote{Likewise, many \emph{Lenia} gliders turn into space-filling patterns when destabilized, though more explosively.}
The sharp cliff in $\mu$, visible in Figs.~\ref{fig:scan1} (B) and~\ref{fig:scan2} (top left, thick blue line), thus provides an overestimate of $r_\ast$, and matches the spontaneous growth threshold of a single cell (see the sequence of dividing patterns in Fig.~\ref{fig:scan2}).
The cliff shifts towards $r_\ast$ as collisions become more frequent (through more seeds or longer simulations).
The edge of growth therefore appears as a collective transition governed by interactions among patterns, for which single-cell growth thresholds provide useful but generally biased estimates.

In the upper basin, the edge has been estimated by bisecting $r$ to locate the value at which a single snake-like pattern neither grows nor shrinks, as shown at the bottom of Fig.~\ref{fig:scan2}. This single-snake estimate is reasonably accurate because snake elongation is the main growth mechanism in this slice of parameter space. However, a snake making a U-turn can still generate growth below the estimated $r$.
Here again, collective effects and self-contact affect the edge of growth, producing a slight misalignment between the estimate (top left, orange line) and the scan.

The lower edge is particularly delicate to estimate across $\lambda$ because it involves a variety of growth-generating patterns.
Those illustrated in Fig.~\ref{fig:scan2} are located on the cliff (top left, thick blue line).
For $\lambda>3$, metastable growing bubbles can occur at relatively low $r$ and even produce numerous gliding cells when destabilized. Those cells are fragile, however, and can readily disappear after collisions.
The statistical nature of the transition is particularly apparent in this regime.

For $\lambda<3$, chains and filaments emerge and can percolate into a foam that persists below the main density cliff. The extent of the foam region (top left, thin blue line) is estimated from a secondary jump, enhanced with 50 seeds instead of 5. The foam is generated by gliding cell-like patterns at the tips of extending filaments. As shown in Figs.~\ref{fig:scan2} and~\ref{fig:edge}, this mode of growth is reminiscent of spores and hyphae forming a mycelium.

\subsection{Self-organization near the edge of growth}

We finally conduct a self-tuned estimation of the lower edge of growth based on the resource-limitation mechanism. Starting from an overestimate $r_1>r_\ast$ at a given $\lambda$, we apply feedback of type II to the anchor point $\mathbf{p}/r_1$. As the total resource $R$ is decreased at fixed volume $|\Omega|=2^{12}$, we monitor the resulting resource-abundance scaling $r(t)=r_1 R_\text{a}(t)/R$.
The resource must be limiting enough to leave the volume-saturated regime (extensivity with $R$ must be verified), but also sufficient to generate many interacting cells.
The abundance at the edge $r_\ast$ is estimated as the equilibrium value of $r(t)$, extracted from a noisy plateau in the suitable $R$-range. Without increasing $|\Omega|$, the estimate can be refined by lowering $r_1>r_\ast$ to enlarge this range.

Disappearance provides an imperfect restoring mechanism from the dilute side, since cells can survive down to the droplet stability boundary. Decreasing $R$ can therefore produce mass hysteresis and underestimate $r_\ast$, unless the system is restarted at a lower mass to ensure that the edge is approached from the dense side.

Filament and chain patterns are most affected by the limited volume, since their dynamics change when they percolate into a foam. In unbounded space, extending filaments are typically destabilized by resource depletion in the course of growth. The global resource coupling leads to a sudden massive collapse, repeated episodically if $r_1$ is sufficient to reignite growth in the system. In this case, $r(t)$ never reaches equilibrium, so a representative intermediate value is reported.

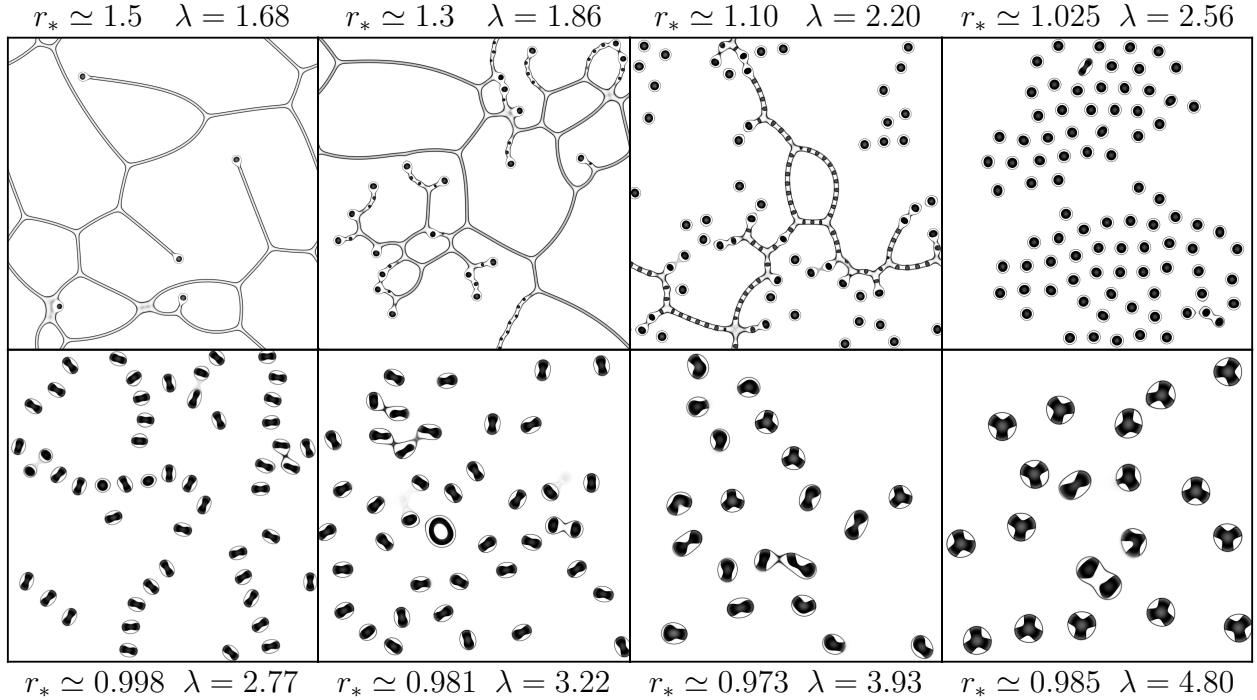
\begin{figure*}[htb]
    \centering
    \vspace{10pt}

    \begin{overpic}[width=\linewidth, decodearray={1 0 1 0 1 0}]{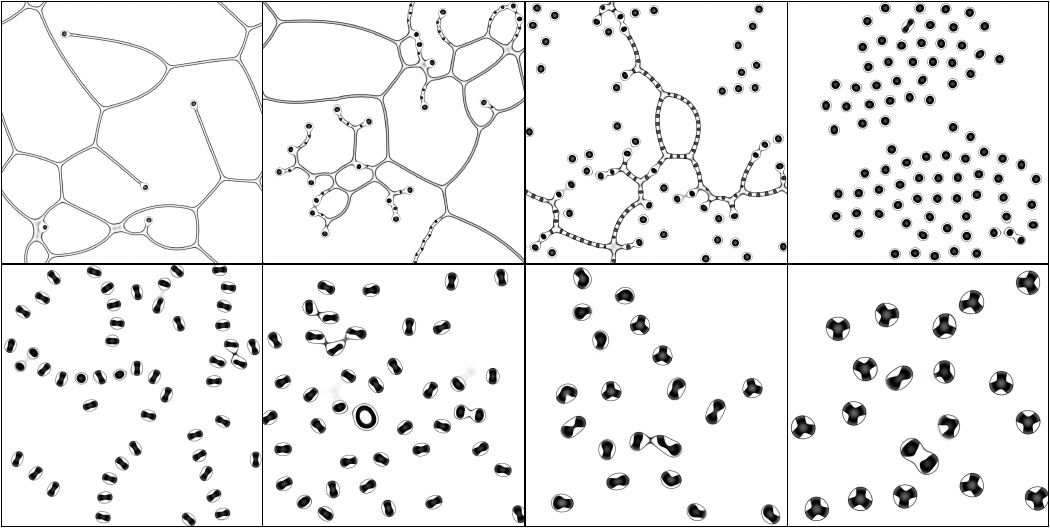}
    \put(1.5,51.2){$\;r_\ast \simeq 1.5 \;\;\;\,\lambda = 1.68$}
    \put(26.5,51.2){$\;r_\ast \simeq 1.3 \;\;\;\,\lambda = 1.86$}
    \put(51.5,51.2){$\,r_\ast \simeq 1.10 \;\;\;\lambda = 2.20$}
    \put(76.5,51.2){$r_\ast \simeq 1.025 \;\;\lambda = 2.56$}
    \put(1.5,-2.2){$r_\ast \simeq 0.998 \;\;\lambda = 2.77$}
    \put(26.5,-2.2){$r_\ast \simeq 0.981 \;\;\lambda = 3.22$}
    \put(51.5,-2.2){$r_\ast \simeq 0.973 \;\;\lambda = 3.93$}
    \put(76.5,-2.2){$r_\ast \simeq 0.985 \;\;\lambda = 4.80$}
    \end{overpic}\\[12pt]
    \caption{Life-like patterns self-organized near the (lower) edge of growth, with resource-limitation feedback of type II and domain volume $|\Omega|=2^{12}$. Their estimated coordinates $(r_\ast,\lambda)$ are indicated by blue markers in Fig.~\ref{fig:scan2}.}
    \label{fig:edge}
\end{figure*}

Figure~\ref{fig:edge} shows patterns generated by these resource-limited estimates, hence near the edge of growth.
Despite fluctuations in mass and available resource, they can be mapped in Fig.~\ref{fig:scan2} (top left, blue markers) with coordinates $(r\simeq r_\ast,\lambda)$.
The delicate balance between growth- and decay-inducing dynamics is encoded in a surprising variety of near-transition morphologies.

\section{Discussion and outlook}
\label{sec:discussion}

\subsubsection*{Synthesis}

Strikingly, translating the \emph{Game of Life} into a continuous and minimal model leads to the emergence of cell-like patterns with a nucleus and a shell, capable of self-replication, motility, and disappearance.
The cGoL model admits a coarse-grained, non-standard reaction--diffusion interpretation: its two auxiliary fields encode the state of the neighbourhood at distinct scales and serve as fast-relaxed morphogen concentrations, produced through a cell-sourced cascade of intermediate reactions.
The nonlinear survival rule then specifies growth conditions as homeostatic concentration ranges for these morphogens.
In this chemical interpretation, the shell is maintained in a narrow band near a low-concentration contour, enclosing a nucleus regulated by a broader range of higher morphogen concentrations.
Slight variations around this morphology reveal a rich diversity of behaviours and dynamical phenomena.
These are organized near a dynamical phase transition separating a dilute phase with quiescent patterns from a dense one in which cells proliferate through successive divisions.
A related transition also appears in a distinct region of parameter space associated with a non-vanishing background state, where snake-like patterns elongate or shrink into droplets.

Assuming that growth locally consumes a diffusive environmental resource, the model can be embedded in a mass-conserving dynamics; in the well-mixed regime, this reduces to a global resource constraint.
Resource limitation then acts as a self-organization mechanism: initialized in the dense phase, the system is effectively retuned by feedback and driven towards the edge of growth.
At this boundary, collective interactions can stimulate divisions below the spontaneous growth threshold of an isolated cell.
Across neighbourhood scale ratios, the same self-tuning procedure estimates the location of the edge and reveals a diversity of surviving and growth-generating life-like morphologies.

\subsubsection*{Dynamical systems}

Our description of the emergent cells is primarily phenomenological.
Beyond the saddle-node structure of homogeneous states, localized morphologies are described in terms of observed symmetry breaking, shape instabilities, bifurcations, and finite-amplitude transitions.
This empirical picture suggests that the reference cell lies near several long-lived localized morphologies and close to a collective dilute-to-dense transition, where self-replication either ceases or is balanced by disappearance.
A natural continuation is to analyse these phenomena with the dynamical-systems tools recently applied to localized patterns in continuous-time \emph{Lenia} by~\citet{Yevenko25}: Lyapunov spectra, covariant Lyapunov vectors, and symmetry-generated neutral modes provide a linear-perturbation framework.
In the present model, this framework could characterize infinitesimal instabilities of isolated cells, distinguish regular from chaotic regimes, and guide the construction of bifurcation diagrams linking eggs, gliders, and related morphologies.

However, some division events, collision-induced switches, and activity avalanches involve finite-amplitude transitions; the associated infinitesimal instabilities can nevertheless be probed by linearizing the dynamics along trajectories in which these events occur.
Self-replication raises an additional challenge, since the number of localized patterns changes in time and the effective attractor dimension may become extensive in the dense phase.
A full dynamical description should therefore connect single-pattern stability with finite-time diagnostics and event-conditioned perturbations, then extend to basin-level and statistical descriptions of collective transitions.

\subsubsection*{Self-organization and criticality}

The resource feedback illustrates how a conservation law, introduced into otherwise non-conserved dynamics, can become an organizing principle: it embeds parameter tuning into the dynamics itself.
This invites comparison with classical self-organized critical systems, while also exposing important differences. The model combines several relevant ingredients: conservation, an autonomous approach to a transition in the large-volume limit, a boundary reminiscent of quiescent-to-active transitions, and robustness to feedback details and parameter variations, provided the system is initialized in the active phase. Important differences remain: the restoring mechanism on the dilute side is imperfect, drive and relaxation are not imposed as separate processes, and the relevant activity order parameter remains unclear in a continuous system where persistent localized dynamics need not imply net growth.

Nevertheless, previous studies have reported evidence for self-organized criticality or near-criticality in the discrete \emph{Game of Life}~\citep{Bak89,Alstrom94,Reia14}; its continuously adjustable logistic extension displays two critical points~\citep{Akgun26}, one with a quiescent-to-active transition resembling the edge of growth, the other percolation-like and perhaps closer to the background-basin change observed here.
In the cGoL, the jump in the density order parameter suggests that parts of the sampled edge behave as first-order or coexistence-like transitions; however, this does not exclude near-critical behaviour, nor a genuine continuous transition in an underlying activity observable tied to proliferation, disappearance or avalanches. More broadly, the present $(r,\lambda)$ scan samples only a restricted slice of a likely high-dimensional edge-of-growth manifold, which already appears to include coexistence fronts, hysteretic foams or filaments with episodic collapses, and long-range organized states such as the polarized cell alignments visible in Fig.~\ref{fig:edge}. This diversity is naturally discussed within the broader framework reviewed by~\citet{Buendia20}, which includes self-organization near criticality, self-organized bistability, and related scenarios. Distinguishing these possibilities will likely require activity measures adapted to continuous life-like patterns, together with avalanche statistics, hysteresis tests, finite-size scaling and correlation diagnostics.

\subsubsection*{Dimension, degrees of freedom and evolution}

The spatial dimension was fixed to $d=2$ for the main presentation, but it remains a structural variable controlling the morphologies and instabilities available to localized patterns.
At fixed parameters, lower-dimensional solutions can be embedded in higher dimensions by extending them uniformly along the additional coordinates: 1-dimensional patterns become stripes or walls, while 2-dimensional cells become columns.
Transverse instabilities can destabilize these embedded states and generate new morphologies.
Conversely, $d=1$ lacks the transverse pinching mechanism required for autonomous scission, although certain metastable growing patterns can split after external perturbations.
Cell division also occurs in $d=3$, and gliders or oscillators occur in both $d=1$ and $d=3$, though generally at shifted parameter values.
Understanding how dimension affects morphology and stability may reveal a more natural parametrization of the target function and further reduce the model's effective degrees of freedom.

The target function was deliberately kept parsimonious, with 6 target parameters and the geometric scale ratio $\lambda$.
Nevertheless, locating a life-like region such as the one presented here would be difficult without relying on successive rule refinements, from the \emph{Game of Life} cellular automaton to the continuous-space system \emph{SmootherLife}~\citep{CornusAmmonis17}, the immediate precursor of the present formulation.
Resource feedback reframes this tuning problem rather than removing it: the abundance parameter, previously hand-tuned and implicit, is replaced by a fixed total resource, an environmental quantity that need not itself be carefully tuned.
This cell--environment split also confers robustness: the system needs only to start in the dense phase, rather than at the edge, since it self-organizes towards it.
The harder challenge remains identifying, for other parametrizations, comparable regions of dilute-to-dense transition that support such self-replicating, life-like behaviour---or discovering novel self-tuning mechanisms that narrow this search.

Finally, viewing the cGoL as a minimal model for hypothetical primordial life-forms raises the question of evolution.
A biologically motivated extension, building on preliminary localized-parameter implementations~\citep{Plantec23,Chan23}, would allow each emergent cell to carry distinct heritable values subject to mutations at the individual level.
Selection would then act indirectly on these parameters through phenotypic behaviours such as self-replication, motility, and robustness to interactions, previously used as explicit fitness criteria~\citep{Davis22b}.
Persistence, propagation, and disappearance---all emergent from the dynamics itself---would thus define an autonomous selection process in which individual mutations drive a coupled stochastic exploration of higher-dimensional rule parameterizations across interacting cells.
Preliminary observations suggest the importance of self-organization near the edge of growth: without negative feedback, selection favours rapid spatial expansion, as also observed in~\citep{Chan23}, whereas resource limitation keeps the dynamics in a marginal regime where survival and reproduction depend on the detailed behaviour of localized patterns.
Implementing such inter-individual diversity and studying its evolutionary consequences will be addressed in a separate article.

\section*{Data and Code Availability}

Code is available at: \href{https://codeberg.org/A-Guillet/cGoL}{codeberg.org/A-Guillet/cGoL}.

\section*{Acknowledgements}

We acknowledge the cross-fertilizing contributions from citizen science, generative arts and software engineering, which have fostered important progress, not only through models but also via efficient cross-platform implementations, interactive tools and empirical exploration techniques. These help manage the complexity of these life-like systems and enable enthusiasts to explore their beauty.
A.G. gratefully acknowledges Théotime Girardot, CornusAmmonis, Daniel M.\ Busiello, Jérémy Sourd, Hiroki Sayama and the anonymous reviewers for their valuable feedback and thoughtful suggestions, which played a significant role at various stages of development of this project.

This work has been funded by the Max Planck Society.

\printbibliography

\end{document}